\documentclass{aastex631}

\graphicspath{{./}{figures/}}

\received{}
\revised{}
\accepted{}
\submitjournal{ApJ Letters}

\shorttitle{Markovian Features of the Solar Wind at Sub-Proton Scales}
\shortauthors{Benella et al.}

\begin{document}
\title{Markovian Features of the Solar Wind at Sub-Proton Scales}

\correspondingauthor{Simone Benella}
\email{simone.benella@inaf.it}

\author[0000-0002-7102-5032]{Simone Benella}
\affiliation{INAF-Istituto di Astrofisica e Planetologia Spaziali, 00133 Roma, Italy}

\author[0000-0002-6303-5329]{Mirko Stumpo}
\affiliation{Dip. Fisica, Università degli Studi di Roma Tor Vergata, 00133 Roma, Italy}
\affiliation{INAF-Istituto di Astrofisica e Planetologia Spaziali, 00133 Roma, Italy}

\author[0000-0002-3403-647X]{Giuseppe Consolini}
\affiliation{INAF-Istituto di Astrofisica e Planetologia Spaziali, 00133 Roma, Italy}

\author[0000-0001-6096-0220]{Tommaso Alberti}
\affiliation{INAF-Istituto di Astrofisica e Planetologia Spaziali, 00133 Roma, Italy}

\author[0000-0002-3182-6679]{Vincenzo Carbone}
\affiliation{Dipartimento di Fisica, Università della Calabria, Rende (CS), 87036, Italy}

\author[0000-0001-5481-4534]{Monica Laurenza}
\affiliation{INAF-Istituto di Astrofisica e Planetologia Spaziali, 00133 Roma, Italy}

\begin{abstract}

The interplanetary magnetic field carried out from the Sun by the solar wind displays fluctuations over a wide range of scales. While at large scales, say at frequencies lower than 0.1-1 Hz, fluctuations display the universal character of fully developed turbulence with a well defined Kolmogorov-like inertial range, the physical and dynamical properties of the small-scale regime as well as their connection with the large-scale ones are still a debated topic. In this work we investigate the near-Sun magnetic field fluctuations at sub-proton scales by analyzing the Markov property of fluctuations and recovering basic information about the nature of the energy transfer across different scales. By evaluating the Kramers-Moyal coefficients we find that fluctuations in the sub-proton range are well described as a Markovian process with Probability Density Functions (PDFs) modeled via a Fokker-Planck (FP) equation. Furthermore, we show that the shape of the PDFs is globally scale-invariant and similar to the one recovered for the stationary solution of the FP equation at different scales. The relevance of our results on the Markovian character of sub-proton scale fluctuations is also discussed in connection with the occurrence of turbulence in this domain.

\end{abstract}


\section{Introduction} \label{sec:intro}

The recently launched Parker Solar Probe (PSP) mission~\citep{fox2016solar} has increased the interest in investigating the evolution of solar wind properties through the inner Heliosphere. Several studies have been performed to investigate the near-Sun or pristine solar wind properties~\citep{Bale19} and to characterize the radial evolution of magnetic field fluctuations at different heliocentric distances in terms of high-order statistics of increments~\citep{Alberti20}, spectral features~\citep{chen2020evolution}, entropic character of magnetic field fluctuations~\citep{stumpo2021self}, the emergence of large-scale rapid polarity reversals known as switchbacks~\citep{Dudok20}. PSP measurements allows not only to investigate the solar wind features at different heliocentric distances, but also to characterize the dynamics of its fluctuations over a wide range of scales, from the large scales up to the sub-proton regime. The inertial range properties, i.e., the behavior of magnetic field fluctuations between the integral scale $L$ and the ion inertial length scale $d_i$, seem to be consistent with expectations from the magnetohydrodynamic (MHD) turbulence picture of describing the solar wind fluctuations in a fluid-like approximation, whereas the behavior of fluctuations in the sub-proton regime, i.e., at scales smaller than $d_i$, still remains unclear and highly debated~\citep{chhiber2021subproton}. 

The characterization of physical processes operating across the inertial range and responsible of transferring energy towards smaller scales has been broadly investigated in terms of stochastic processes. This is a long-standing idea since pioneering works by \citet{ruelle71on} and \citet{mandelbrot78geometrical}. In the framework of hydrodynamic turbulence it has been shown that the statistics of longitudinal velocity increments can be described in terms of Markov \textit{process in scale} ~\citep{pedrizzetti1994markov,friedrich1997statistical,davoudi1999theoretical,renner2001experimental}. Indeed, the main idea behind these works is to represent the statistics of the longitudinal velocity increments as a stochastic process evolving across the length or time scales, instead of the common evolution in space or in time. A similar approach aiming to study the markovian properties of the solar wind magnetic field fluctuations in the inertial range, has been used by ~\citet{strumik2008statistical,strumik2008testing} in the framework of space plasma turbulence. They showed that the turbulent cascade in the solar wind satisfies the Markov condition, suggesting the presence of a local energy transfer mechanism between subsequent scales which therefore does not depend on large-scale structures or the driving mechanisms of solar wind turbulence. However, this kind of study for the sub-proton regime is still missing. One of the striking features of this regime is the existence of a scale-invariant nature suggesting a filamentary structure of the dissipation field~\citep{Alberti21}. It has been also interpreted as the existence of a scale-invariant topology of current sheets between ion and electron inertial scales~\citep{chhiber2021subproton}, leading the system towards a restored symmetry of the statistics of fluctuations~\citep{Dubrulle19}. Nevertheless, there is no general consensus on the physical mechanisms explaining these small-scale features, which are generally highlighted through spacecraft measurements, while are not observed through numerical simulations~\citep{Papini19,Papini21}. 

The properties of the magnetic field fluctuation statistics, expressed in terms of the scaling of power spectrum and structure functions, display a universal character in the inertial range of plasma turbulence. Conversely, in the sub-proton range a deep understanding of these statistical properties is still missing. In this paper we investigate the Markovian character of the magnetic field fluctuations at sub-proton scales by using high-resolution measurements gathered by PSP in the near-Sun solar wind for the first time. We provide a parameterization of the Kramers-Moyal coefficients associated with the magnetic field fluctuations as a function of the time scale. We show that the time-scale evolution of the probability density functions (PDFs) is governed by the Fokker-Planck (FP) equation. As already shown in previous works, the shape of the experimental PDFs exhibits a global scale-invariance. In this work we show that these pdfs can be successfully approximated by considering the stationary solution of the corresponding FP equation, contrary to what is generally observed in the inertial range of fully developed turbulence.

\section{Data Description and Methods\label{sc1}}
\subsection{Data}

We investigate the statistics of the magnetic field increments at sub-proton scales in the pristine solar wind by using measurements gathered by the FIELDS suite on board PSP \citep{bale2016fields}. Specifically, we focus on a 28-minutes interval on 2018-11-06 from 02:20:00 UT to 02:48:00 UT, when PSP was located at about 0.17 AU from the Sun. We use data from the SCaM data product in the spacecraft reference frame, which merges measurements from the fluxgate (FGM) and the search-coil (SCM) magnetometers, enabling observations from DC up to 1 MHz with an optimal signal-to-noise ratio \citep{bowen2020merged}. Here, we consider data with a sampling of 293 samples/s, corresponding to $\sim 0.0034$ s time resolution.

Since the spacecraft reference frame has no particular significance for our results, we rotate the magnetic field components in the minimum variance reference system, where $B_1$, $B_2$ and $B_3$ are the minimum, intermediate and maximum variance components, respectively. The minimum variance component mainly resides along the radial direction while intermediate and maximum variance components are representative of fluctuations along the transverse directions with respect to the mean field.

 \begin{figure*}
  \vskip 0.5 cm
  \centering
 	\includegraphics[width=16cm]{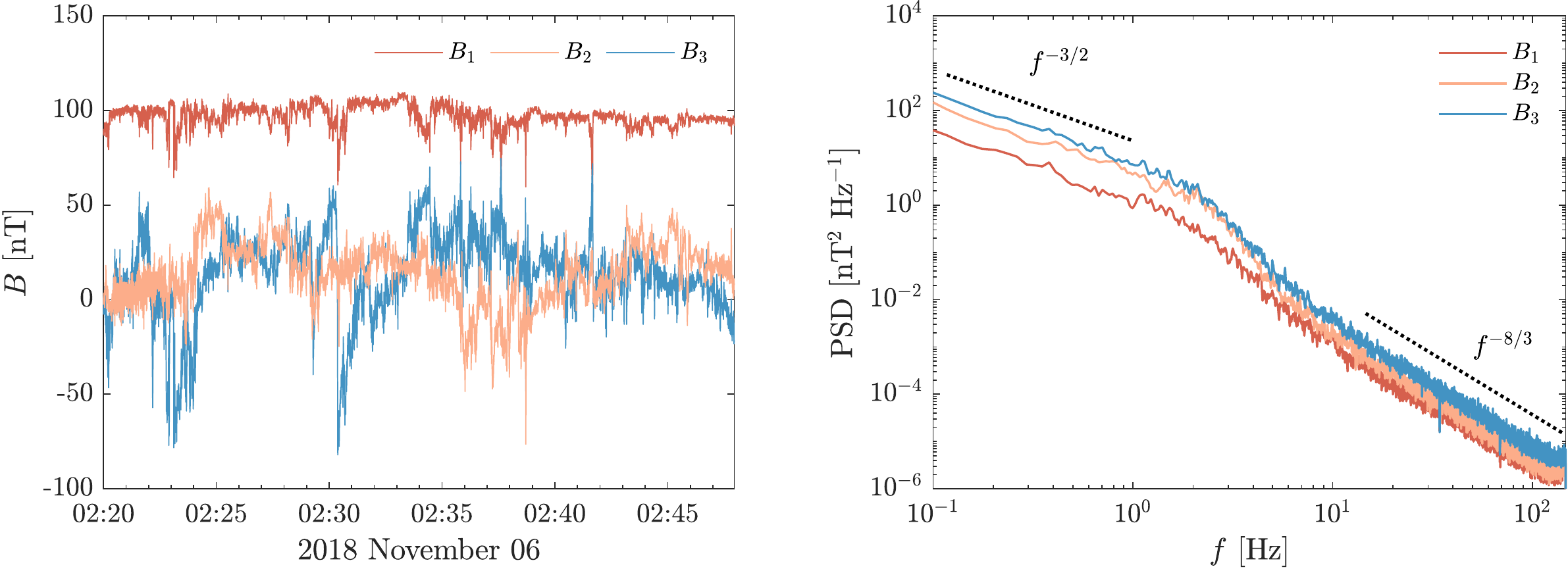}
	 \caption{Left panel: the time series of the minimum ($B_1$), intermediate ($B_2$) and maximum ($B_3$) variance components of the magnetic field. Right panel: the corresponding PSDs. Dashed lines refer to power-law trends characterized by spectral slopes -3/2 at low frequencies and -8/3 at high frequencies, respectively. \label{fig:b-psd}}
 \end{figure*}

Figure \ref{fig:b-psd} shows the time series of the magnetic field components (left panel) along with the corresponding Power Spectral Densities (PSDs, right panel). As usually observed the PSDs show two different spectral behaviors $f^{-\beta}$: at low frequencies (i.e., $f \lesssim 1$ Hz) $\beta \in [3/2, 5/3]$ \citep[in agreement with recent findings,][]{Alberti21,chen2020evolution,chhiber2021subproton}, while in the sub-proton domain (i.e., $f \gtrsim 10$ Hz) $\beta \in [7/3,8/3]$ \citep{chhiber2021subproton}. Moreover, a transition region is also observed between the two different dynamical regimes, with exponents showing a transition from inertial to sub-proton range behavior. 

\subsection{Methods}


A fundamental quantity in the analysis of solar wind turbulence is represented by the magnetic field increment (fluctuation) across a time separation scale $\tau$, defined as
\begin{equation}
    b_{i, \tau} \doteq B_i(t+\tau) - B_i(t),\quad i=1,2,3\ .
    \label{eq:incr}
\end{equation}
These increments represent a stochastic process in $\tau$, thus it is relevant to investigate their Markovian character. Given a stochastic process $x(t, \tau)$, the main quantity in defining a Markov process is the transition probability, i.e., the probability of observing the state $x_1$ at the scale $\tau_1$ given the states $x_2$ at the scale $\tau_2$ until $x_n$ at the scale $\tau_n$, with $\tau_1<\tau_2<\dots<\tau_n$. The process is Markovian if the $n$-point transition probability satisfies the condition
\begin{equation}
    p(x_1,\tau_1|x_2,\tau_2;\dots;x_n,\tau_n)=p(x_1,\tau_1|x_2,\tau_2),
    \label{eq:markov}
\end{equation}
and then the knowledge of the initial distribution $p(x_n,\tau_n)$ and the two-points transition probabilities allows a complete knowledge of $n$-point probability distribution.

An important relation arising from the Markov condition is the Chapman-Kolmogorov (CK) equation expressing the transition probability of observing $x_1$ at the scale $\tau_1$ given $x_3$ at the scale $\tau_3$ by integrating on a variable $x_2$ at an arbitrary intermediate scale $\tau_1<\tau_2<\tau_3$ \citep[][p. 28]{renner2001experimental,risken1996fokker}, i.e., 
\begin{equation}
    p(x_1,\tau_1|x_3,\tau_3)=\int_{-\infty}^{+\infty} p(x_1,\tau_1|x_2,\tau_2)p(x_2,\tau_2|x_3,\tau_3)dx_2.
    \label{eq:ck}
\end{equation}

The differential form of the CK equation is called \textit{master equation} and reads as
\begin{equation}
   -\frac{\partial}{\partial \tau}p(x,\tau|x',\tau')=\mathcal{L}_{KM}(x,\tau)p(x,\tau|x',\tau').
    \label{eq:master}
\end{equation}
Equation (\ref{eq:master}) expresses the time evolution of the transition probability in terms of the Kramers-Moyal (KM) expansion and the minus sign on the left-hand side of Equation (\ref{eq:master}) is due to the direction of the time evolution towards smaller scales \citep{renner2001experimental}. Here, the operator $\mathcal{L}_{KM}(x)$ is the KM operator, 
\begin{equation}
    \mathcal{L}_{KM}(x,\tau)=\sum_{k=1}^\infty\biggl(-\frac{\partial}{\partial x} \biggr)^k D^{(k)}(x,\tau),
\end{equation}
where the functions $D^{(k)}(x,\tau)$ are the KM coefficients. 

For a generic stochastic Markov process all the terms in the KM expansion are different from zero. However, according to the Pawula's theorem, if the fourth-order coefficient $D^{(4)}(x,\tau)$ is equal to zero, all the coefficients of order $k\geq3$ vanish and the KM expansion stops at the second-order. In that case, the KM expansion reduces to the Fokker-Planck (FP) equation \citep{risken1996fokker}
\begin{equation}
  	 -\frac{\partial}{\partial \tau}p(x,\tau|x',\tau')
   	 =\biggl[-\frac{\partial}{\partial x}D^{(1)}(x,\tau)+\frac{\partial^2}{\partial x^2}D^{(2)}(x,\tau)\biggr]p(x,\tau|x',\tau').
    \label{eq:fp}
\end{equation}
The first-order KM coefficient $D^{(1)}(x,\tau)$ represents the \textit{drift} function, accounting for the deterministic evolution of the stochastic process $x$, whereas the second-order KM coefficient $D^{(2)}(x,\tau)$ constitutes the \textit{diffusion} term, which modulates the amplitude of the delta-correlated Gaussian noise, $\Gamma(\tau)$, of the corresponding Langevin equation
\begin{equation}
    -\frac{\partial x}{\partial\tau}=D^{(1)}(x,\tau)+\sqrt{D^{(2)}(x,\tau)}\,\Gamma(\tau).
\end{equation}

From a practical point of view, the KM coefficients are not directly accessible from data, but rather they can be evaluated by using the conditional moments. According to \cite{renner2001experimental,renner2001evidence}, the $k^{th}$-order conditional moment is defined as
\begin{equation}
    M^{(k)}_{\Delta\tau}(x,\tau)=\int_{-\infty}^{+\infty}(x'-x)^k p(x',\tau+\Delta\tau|x,\tau)dx'.
    \label{eq:moments}
\end{equation}
The corresponding $k^{th}$-order KM coefficient is defined by taking the limit
\begin{equation}
    D^{(k)}(x,\tau)=\frac{1}{k!}\,\lim_{\Delta\tau\to0}\,\frac{1}{\Delta\tau}\,M^{(k)}_{\Delta\tau}(x,\tau).
    \label{eq:kmcoef}
\end{equation}
Whereas conditional moments defined in (\ref{eq:moments}) can be computed from the experimental observations, the definition (\ref{eq:kmcoef}) cannot be applied exactly. In fact, the best estimate of the $k^{th}$-order KM coefficient considered in the analysis is given by
\begin{equation}
    D^{(k)}_{\tau_s}(x,\tau)=\frac{1}{k!\tau_s}\,M^{(k)}_{\tau_s}(x,\tau),
    \label{eq:km-res}
\end{equation}
where $\tau_s$ indicates the time resolution of the time series.

Here, we applied the above analysis to the small-scale increments of the magnetic field at the sub-proton scales, starting from the verification of the CK equation (\ref{eq:ck}), and successively evaluating the KM coefficients up to the fourth order.

\section{Results\label{sc2}}

\begin{figure*}
\centering
 	\includegraphics[width=15 cm]{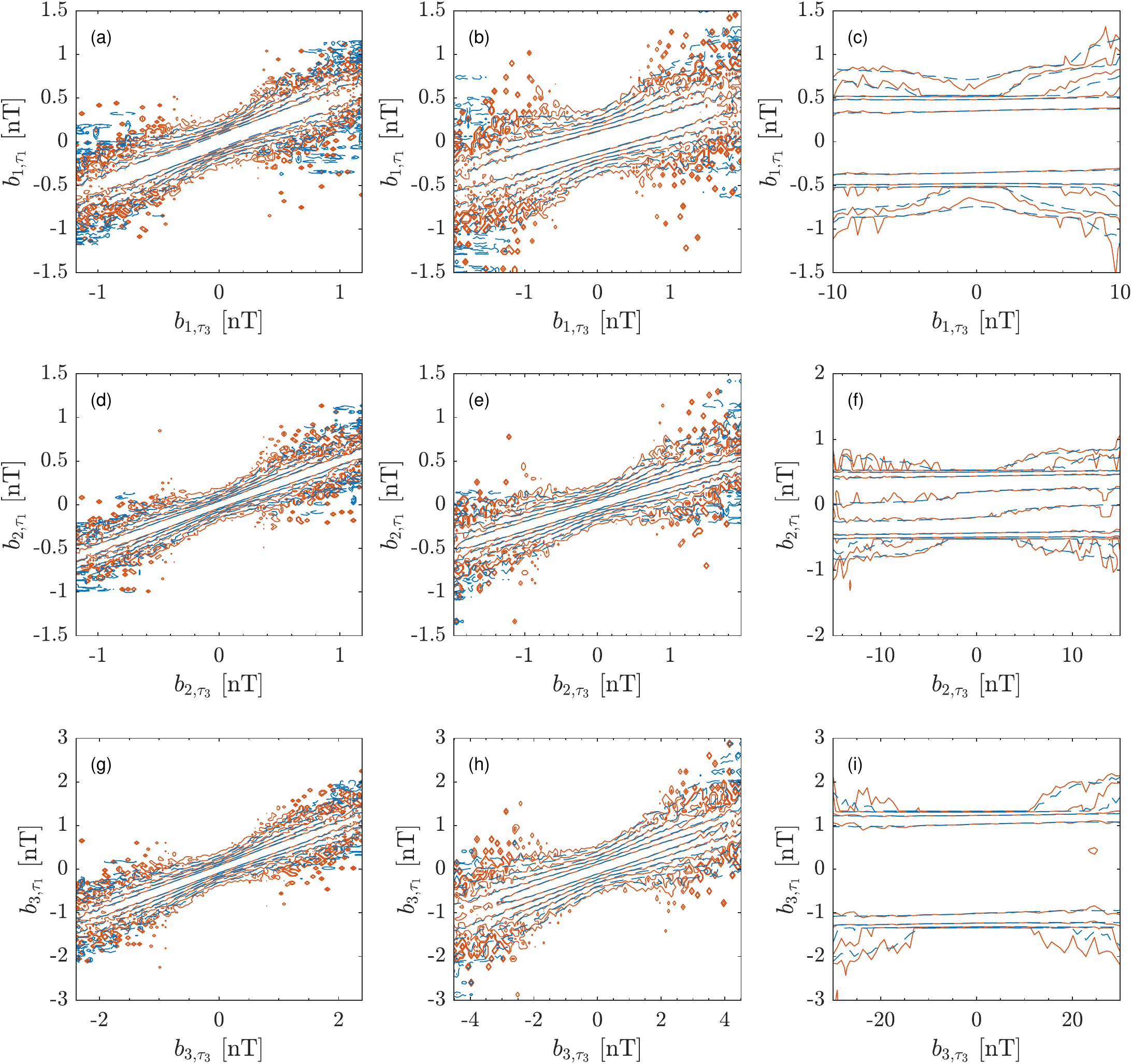}
	 \caption{Comparison between observed (red curves) and reconstructed (blue curves) conditional probabilities at different time scales for minimum (upper panels), intermediate (middle panels) and maximum (lower panels) variance directions. The time-scale differences $2\,\Delta\tau=\tau_3-\tau_1$ for the CK test are: 0.0068 s (panels a, d and g), 0.02 s (panels b, e and h) and 1.0 s (panels c, f and i).\label{fig:ck-test}}
\end{figure*}

The first step of our analysis consists in searching for the Markovian nature of the statistics of increments $b_{i, \tau}$. Although the analysis is performed in the temporal domain, it is possible to assume that we are exploring spatial scales via Taylor's hypothesis, i.e., $r = \tau V_{SW}$ where $V_{SW}$ is the solar wind velocity. This hypothesis has been shown to be valid in the selected time interval being solar wind supersonic and super-Alfv\'enic \citep[e.g.,][]{chhiber2021subproton,perez2021applicability}. In the following we show the results for the minimum, intermediate and maximum variance directions.
One way to test the Markov condition on the statistics of increments is to prove that Equation (\ref{eq:ck}) is satisfied for the stochastic process defined by the increments $b_{i,\tau}$ at different time scales. In the following, we set the small time scale $\tau_1=0.01$ s and we show results for three different values of the time scale separation $\Delta\tau=0.0034$, 0.01 and 0.5 s. The intermediate time scale is then evaluated as $\tau_2=\tau_1+\Delta\tau$, while the large time scale as $\tau_3=\tau_1+2\Delta\tau$. The statistics of increments at these scales allows us to evaluate and compare both members of Equation (\ref{eq:ck}). We refer to the left-hand side of Equation (\ref{eq:ck}) as the \textit{empirical} conditional probability, $p_E$, and to the right-hand side as the CK conditional probability $p_{CK}$. The results of the CK test are shown in Figure \ref{fig:ck-test}.
Figures \ref{fig:ck-test}a, \ref{fig:ck-test}d and \ref{fig:ck-test}g display a good agreement between $p_E$ (red lines) and $p_{CK}$ (blue lines) for $\Delta\tau=0.0034$ s, suggesting that the process is markovian at this scale. By increasing the separation to $\Delta\tau = 0.01$ s the Markov condition is still fulfilled in the whole sub-proton range, Figures \ref{fig:ck-test}b, \ref{fig:ck-test}e and \ref{fig:ck-test}h. 
However, by increasing the separation to $\Delta\tau = 0.5$ s, such to fall at the end of the inertial regime, the CK equation appears still valid, although neither $p_E$ nor $p_{CK}$  depends on the large scale increments $b_{i,\tau_3}$, Figures \ref{fig:ck-test}c, \ref{fig:ck-test}f and \ref{fig:ck-test}i. 
Summarizing, for the considered data sample the scale-to-scale process defined by $b_{i,\tau}$ is in general markovian across the whole sub-proton domain and the fluctuation amplitudes in the sub-proton domain seem to be statistically independent from those observed in the inertial range, i.e., $p(b_{i,\tau_1},\tau_1|b_{i,\tau_3},\tau_3)=p(b_{i,\tau_1},\tau_1)$. We emphasize that the same results have been obtained for all the values of $\tau_1$ within the sub-proton range (not shown). 

\begin{figure*}
    \vskip 0.5 cm
    \centering
    \includegraphics[width=16 cm]{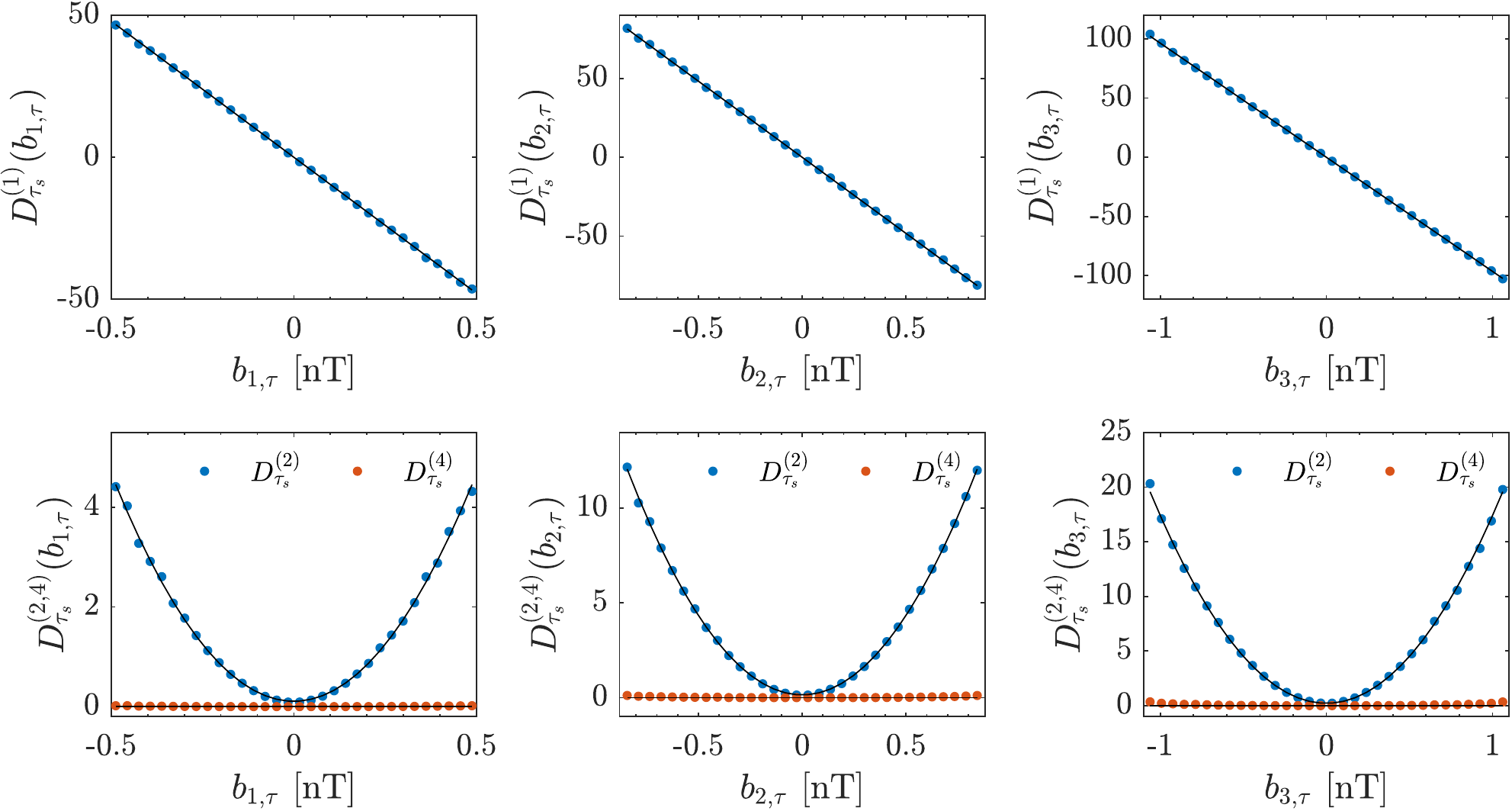}
    \caption{First, second and fourth order finite-size KM coefficients for $b_{1,\tau}$ (left), $b_{2,\tau}$ (center) and $b_{3,\tau}$ (right) at scale $\tau=0.01$ s. Solid lines represent the $D^{(1,2)}_{\tau_s}(b_{i,\tau})$ best fit and $D^{(4)}_{\tau_s}(b_{i,\tau})=0$.
    \label{fig:km-coef}}
\end{figure*}

Previous studies showed that the Pawula's theorem holds across the inertial range of turbulence. This implies that the evolution of the PDFs of field increments in the inertial domain is governed by Equation (\ref{eq:fp}). This result has been accurately validated in the case of hydrodynamic and solar wind turbulence \citep[cf.,][and references therein]{renner2001experimental,peinke2019fokker, strumik2008testing}, while there is no evidence yet at sub-proton scales. Thus, we firstly assess its validity by computing the finite-time scale KM coefficients of the process $b_{i,\tau}$ at the sampling time $\tau_s$, Equation (\ref{eq:km-res}). Figure \ref{fig:km-coef} shows that the fourth order KM coefficient is close to zero for both components suggesting that the Pawula's theorem is also valid in the sub-proton domain. Hence, the evolution of the PDFs of magnetic field increments is governed by the FP equation. Furthermore, it is evident that the first order coefficient is a linear function of $b_{i,\tau}$, whereas the second order coefficient shows a quadratic trend. Thus, we can introduce the following parameterization for $D^{(1)}_{\tau_s}$ and $D^{(2)}_{\tau_s}$ as
\begin{eqnarray}
    D^{(1)}_{\tau_s}(b_i,\tau) = -\gamma_i(\tau) b_i, \label{eq:km0} \\
    D^{(2)}_{\tau_s}(b_i,\tau) = \alpha_i(\tau) + \beta_i(\tau) b_i^2,
    \label{eq:km1}
\end{eqnarray}
such that we can study the dependence of the parameters $\gamma_i(\tau)$, $\alpha_i(\tau)$ and $\beta_i(\tau)$ upon the time scale $\tau$.
All parameters exhibit a power-law dependence on the time scale $\tau$, i.e.
\begin{equation}
    \{\alpha_i(\tau), \beta_i(\tau), \gamma_i(\tau)\} = A_0 \tau^\mu,
    \label{eq:params}
\end{equation}
and the values of $A_0$ and $\mu$ are listed in Table \ref{tab:param} for the three magnetic field components.
\begin{deluxetable}{l|cc|cc|cc}[b!]\bf 
\tablecaption{Fitted coefficients $A_0$ and exponents $\mu$ of Equation (\ref{eq:params}) for $b_{1,\tau}$, $b_{2,\tau}$ and $b_{3,\tau}$ with 95\% confidence bounds.
\label{tab:param}}
\tablecolumns{7}
\tablenum{1}
\tablewidth{0pt}
\tablehead{
\colhead{} &
\multicolumn{2}{c}{$b_{1,\tau}$} &
\multicolumn{2}{c}{$b_{2,\tau}$} &
\multicolumn{2}{c}{$b_{3,\tau}$}}
\startdata
& $A_0$ & $\mu$ & $A_0$ & $\mu$ & $A_0$ & $\mu$\\
\hline
$\alpha$ & $0.7\pm0.2$ & $0.36\pm0.08$
& $1.6\pm0.3$ & $0.47\pm0.06$ & $5.3\pm1.7$ & $0.61\pm0.08$\\
$\beta$ & $0.003\pm0.001$ & $-1.87\pm0.01$
& $0.002\pm0.001$ & $-1.96\pm0.01$ & $0.002\pm0.001$ & $-1.98\pm0.01$\\
$\gamma$ & $0.87\pm0.03$ & $-1.01\pm0.01$
& $0.90\pm0.02$ & $-1.01\pm0.01$ & $0.89\pm0.02$ & $-1.02\pm0.01$\\
\hline
\enddata
\end{deluxetable}

\begin{figure*}
 \vskip 0.5 cm
  \centering
   	\includegraphics[width=16 cm]{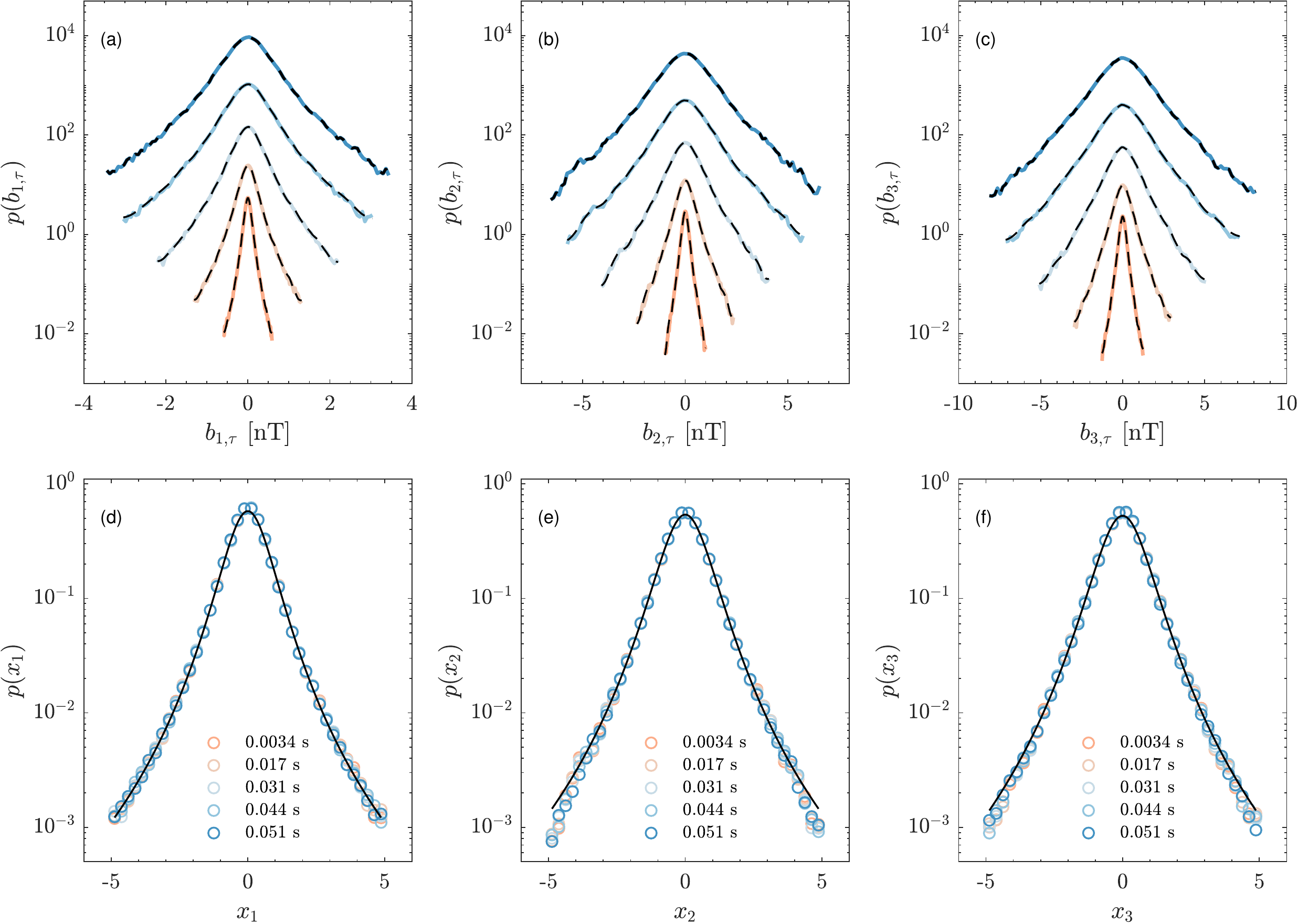}
	 \caption{Comparison of the empirical PDFs $p(b_{i,\tau})$ obtained from the PSP data at different scales (colored lines) with the numerical solutions (dashed lines), panels a, b and c. All the curves are shifted in the vertical direction for clarity of presentation and correspond to the following time scales, from top to bottom: 0.051 s, 0.044 s, 0.031 s, 0.017 and 0.0068 s. Panels d, e and f show the data collapsing of empirical PDFs, circles, along with the best fit of the stationary solution (\ref{eq:kappa}), solid lines. \label{fig:4}}
 \end{figure*}

As a further step, we perform a consistency check by computing the evolution of the PDFs of the magnetic field increments from large towards small scales in the sub-proton domain. We start from $p(b_{i,\tau_0})$ at $\tau_0=0.051$ s as initial condition, and then we compute the numerical solution of the FP equation by assuming a Gaussian short-time propagator \citep{renner2001evidence}.
Figures \ref{fig:4}a, \ref{fig:4}b and \ref{fig:4}c show the comparison between the empirical PDFs and the corresponding FP numerical solutions. The excellent agreement between the empirical PDFs and the theoretical predictions proves that the FP equation with the KM coefficients (\ref{eq:km0}) and (\ref{eq:km1}) accurately describes the evolution of PDFs in the sub-proton domain.

Furthermore, by re-scaling the PDFs at the different scales according to the following transformations, (e.g., see Figures \ref{fig:4}d, \ref{fig:4}e and \ref{fig:4}f, circles),
\begin{eqnarray}
   	 b_{i,\tau}\ \longrightarrow  \ x_i\equiv \frac{b_{i,\tau}}{\sigma_{b_{i,\tau}}},\\
	 p(b_{i,\tau}) \ \longrightarrow  \ p(x_i) \equiv \sigma_{b_{i,\tau}} p( b_{i,\tau})
\label{eq:scale-transf}
\end{eqnarray}
where $\sigma_{b_{i,\tau}}$ is the standard deviation of $b_{i,\tau}$, we obtained a PDF collapsing in the sub-proton range \citep{Kiyani09,Osman_2015,chhiber2021subproton}. The PDF collapsing defines a \textit{master curve} for the shape of the PDFs. Thus, we attempt a comparison between the experimental PDFs of the rescaled increments and the corresponding stationary solutions of the FP equation. We evaluate the stationary distribution $p_{ST}(x_i)$ by solving the time-scale independent FP equation \citep{risken1996fokker}, i.e.,
\begin{equation}
    \frac{\partial}{\partial x_i}[D^{(2)}(x_i)p_{ST}(x_i)]=\frac{D^{(1)}(x_i)}{D^{(2)}(x_i)}D^{(2)}(x_i)p_{ST}(x_i).
    \label{eq:FP-stat}
\end{equation}
By introducing the set of KM coefficients $D^{(1)}(x_i)=-\gamma_i x_i$ and $D^{(2)}(x_i)=\alpha_i+\beta_i x_i^2$ in (\ref{eq:FP-stat}), where $\alpha_i$, $\beta_i$ and $\gamma_i$ are now constants, the stationary solution reads
\begin{equation}
    p_{ST}(x_i) =N_0(\alpha_i+\beta_i x_i^2)^{-\frac{\gamma_i}{2\beta_i}-1}\ ,
\end{equation}
where $N_0$ is the normalization factor.
We may note how the obtained function is a \textit{Kappa distribution} \citep{Milovanov2000,Leubner2005}, that can be written also in the standard form, 
\begin{equation}
	p_{ST}(x_i)  = N'_0 \left[1+\frac{1}{\kappa}\frac{x_i^2}{x_0^2}\right]^{-\kappa}\ ,
	\label{eq:kappa}
\end{equation}
where $\kappa = 1+\gamma_i/2\beta_i$, $x_0^2 = 2\alpha_i/(\gamma_i+2\beta_i)$ and $N'_0 = N_0\alpha_i^{-\kappa}$.

Whereas the inertial range is characterized by strongly intermittent magnetic field fluctuations reflecting in a well-known modifications of the PDF-shape moving from the scale of the forcing towards the dissipation scale, the sub-proton range statistics exhibits a global-scale invariance that manifests in the existence of a \textit{shape-invariant} master curve.
The comparison between the empirical PDFs of the normalized variables $x_i$ and the stationary PDFs are reported in Figures \ref{fig:4}d, \ref{fig:4}e and \ref{fig:4}f. The values of the parameters of Equation (\ref{eq:kappa}) obtained by fitting the distributions are $N_0'=0.58$, $x_0=0.75$ and $\kappa=2.0$ for $x_1$, $N_0'=0.54$, $x_0=0.83$ and $\kappa=2.1$ for $x_2$ and $N_0'=0.53$, $x_0=0.85$ and $\kappa=2.1$ for $x_3$. We stress that in this framework the FP equation describes the evolution of PDFs across time scales instead of time, and thus the concept of stationarity has to be intended as an invariance of the rescaled magnetic field increment PDFs for any time scale in the sub-proton range. The agreement between the observed PDFs and Equation (\ref{eq:kappa}) is remarkably good. However, we point out that this stationary distribution has to be considered as a valid approximation of the core of the empirical PDFs in a restricted range of variability of $x_i$ (e.g., within $\pm5\sigma$ as shown in the bottom panel of Figure \ref{fig:4}). Indeed, in this figure is evident that the tails of the distributions display a slight departure from the  stationary solution, especially for $x_2$ and $x_3$, and for increasing values of $x_i$ the tails of the empirical PDFs decrease more rapidly than the tails of the Kappa distribution. 

\section{Discussion and Conclusions \label{sc4}}
 
In this work we have investigated the Markovian character of the magnetic field increments (fluctuations) at sub-proton scales in turbulent near-Sun solar wind. The results clearly evidenced that at these scale, as already observed in the inertial range \citep{strumik2008statistical,strumik2008testing}, the statistics of the magnetic field fluctuations are Markovian and the dynamics along the different scales can be described in terms of a FP equation.
Since universality based on some turbulent-like approach to fluctuations is lost at small scales, different approaches based on some basic concepts of non-equilibrium statistical mechanics, such as the one presented here, could successfully reveal some universal characteristics of the underlying processes which generate fluctuations at these scales \citep{Carbone_2022}.

On a physical side we obtain that the energy transfer has a local character. Although this feature is also present in the inertial range of the solar wind, our analysis shows that no statistical correspondence is found between magnetic field fluctuations observed in inertial and sub-proton regimes. The observed statistical independence does not mean that there is no energy flux from the inertial domain towards the sub-proton scales, but suggests that the mechanisms at the origin of the fluctuations observed in the two regimes are different. 
Moreover, in various recent and past works it was found that the scaling of structure functions of magnetic field increments at sub-proton scales suggests the occurrence of a \textit{global scale-invariance}, i.e., lack of intermittency. This is here clearly supported by the existence of a scale-invariant shape of the PDFs of magnetic field increments, whose shape is well approximated by the stationary solution of the FP equations. Furthermore, we show that by using the common linear and quadratic parametrizations for the first and second KM coefficients, respectively, the scale-invariant distribution coming from the stationary FP equation is the Kappa distribution (\ref{eq:kappa}).
Since $\kappa$ depends on $\gamma_i$ and $\beta_i$, the KM coefficients can be related to the supposed non-extensive character of fluctuations. Indeed, considering Equation (\ref{eq:kappa}) and reminding its relation between $\kappa$ and the non-estensivity parameter $q$ of Tsallis' entropy, i.e., $\kappa = 1/(1-q)$, from the relation $q=\gamma_i/(2\beta_i+\gamma_i)$ we obtain $q\sim0.5$ for the three magnetic field components. These values do not differ significantly from those measured by \citet{Leubner2005} in the low end of the inertial range of solar wind magnetic field fluctuations.

As a last point we would like to discuss the physical consequences of the observed global scale-invariance. As already mentioned above, the emergence of a global scale-invariance at sub-proton scales means that we do not observe intermittency in this regime. This can be linked to the fact that the statistics of the magnetic field increments can be successfully approximated by the stationary solution of the FP equation with respect to the time scale variable $\tau$. This is equivalent to assume that the probability flux is constant across the scales, i.e., $\partial p/\partial\tau\to0$ and $p\to p_{ST}$. From a speculative point of view, the absence of intermittency could indicate that the scaling properties in this regime could be more realistically related to the formation of a self-similar current structure, which is an expected dissipative structure. In other words, the cascade mechanism in the inertial range ends with the formation of a topological structure that is a complex fractal representative, de-facto, the dissipative pattern. Thus, the observed scale invariance could be the counterpart of the fractal topology of the current structure. Clearly, this is a speculative point of view, which, however, could represent the starting point for successive investigation and analysis. An alternative scenario could be the occurrence of wave turbulence in the sub-proton range such that the cascade mechanism may be better described in terms of an energy transfer flow towards the electron scales where heating and dissipation might occur.

In conclusion, here we have provided a strong evidence for the Markovian character of the small scale magnetic field fluctuations at sub-proton scales along with other strong evidences of the global-scale invariant character of these fluctuations on the considered data sample. These results represent a possible indication for finding universal features in the magnetic field fluctuation statistics at sub-proton scales in terms of Markov processes. Further work is necessary in order to unveil similar statistical properties in different samples of high-frequency interplanetary magnetic field observations.

\section*{Acknowledgements \label{sc5}}

The data used in this study are available at the NASA Space Physics Data Facility (SPDF), \url{https://spdf.gsfc.nasa.gov/index.html}. The authors acknowledge the contributions of the FIELDS team to the Parker Solar Probe mission. This work is funded by the Italian MIUR-PRIN grant 2017APKP7T on “Circumterrestrial Environment: Impact of Sun-Earth Interaction”. M.S. acknowledges the PhD course in Astronomy, Astrophysics and Space Science of the University of Rome “Sapienza”, University of Rome “Tor Vergata” and Italian National Institute for Astrophysics (INAF), Italy.

\bibliography{Benellaetal.bib}

\end{document}